# Characterization of the second- and third-harmonic optical susceptibilities of atomically thin tungsten diselenide

Henrique G. Rosa[1,*], Ho Yi Wei[1,2,3,*], Ivan Verzhbitskiy[1,3], Manuel J. F. L. Rodrigues[1,4], Takashi Taniguchi[5], Kenji Watanabe[5], Goki Eda[1,3,6], Vitor M. Pereira[1,3,†], José C. V. Gomes[1,3,‡]

[1]Centre for Advanced 2D Materials (CA2DM), National University of Singapore, 6 Science Drive 2, Singapore 117546

[2]NUS Graduate School for Integrative Sciences and Engineering (NGS), Centre for Life Sciences (CeLS), 28 Medical Drive, Singapore 117456

[3]Department of Physics, National University of Singapore, 2 Science Drive 3, Singapore 117551

[4]Center of Physics and Department of Physics, Universidade do Minho, 4710-057, Braga, Portugal

[5]National Institute for Materials Science, 1-1 Namiki, Tsukuba 305-0044, Japan

[6]Department of Chemistry, National University of Singapore, 3 Science Drive 3, Singapore 117543

*These authors contributed equally to this work

†Corresponding author email: vpereira@nus.edu.sg

‡Corresponding author email: phyvjc@nus.edu.sg

**Abstract:** We report the first detailed characterization of the sheet third-harmonic optical susceptibility, $\chi^{(3)}{}_s$, of tungsten diselenide ($WSe_2$). With a home-built confocal microscope setup developed to study harmonics generation, we map the second and third-harmonic intensities as a function of position in the sample, pump power and polarization angle, for single- and few-layers flakes of $WSe_2$. We register a value of $|\chi^{(3)}{}_s| \approx 0.9 \times 10^{-28}$ m$^3$ V$^{-2}$ at a fundamental excitation frequency of $\hbar\omega = 0.8$ eV, which is comparable in magnitude to the third-harmonic susceptibility of other group-VI transition metal dichalcogenides. The simultaneously recorded second-harmonic susceptibility is found to be $|\chi^{(2)}{}_s| \approx 0.7 \times 10^{-19}$ m$^2$ V$^{-1}$ in very good agreement on the order of magnitude with recent reports for $WSe_2$, which asserts the robustness of our values for $|\chi^{(3)}{}_s|$.

# Introduction

Atomically thin two-dimensional crystals of semiconducting transition-metal dichalcogenides (TMD) are currently a subject of intense research. The most prominent members of this family have been molybdenum disulfide ($MoS_2$) and diselenide ($MoSe_2$), as well as tungsten disulfide ($WS_2$) and diselenide ($WSe_2$). Their unique electronic, optical and structural characteristics are under active scrutiny for applications in photonics, optoelectronics and electronic devices. In addition to the ability of tuning their carrier densities on demand by field-effect, such properties include high charge-carrier mobility[1,2], a direct band gap in monolayer crystals that evolves to indirect with additional layers[3], photoluminescent emission in the visible-NIR spectral range[4–6], high nonlinear optical susceptibilities[7,8], strong spin-orbit coupling[9,10] and spin-valley locking and novel valleytronics phenomena[11].

Owing to their semiconducting character and favorable band gaps for conventional optoelectronics[12], the pressing need for detailed characterization of their intrinsic optical response has stimulated a steady and broad array of experimental results[12,13]. In particular, nonlinear optical applications hinge upon the nature of the second and third-order optical susceptibilities ($\chi^{(2)}$ and $\chi^{(3)}$, respectively), which are conventionally obtained from harmonic generation experiments. Whereas second-harmonic generation (SHG) in the most prominent TMD had been studied in a number of recent experiments[7,8,14–16], reports on third-harmonic generation (THG) among this family have been scarce[17,18]. Remarkably, THG in either single or few-layer selenide-TMD remains unexplored. These results on harmonic generation in TMD and in other 2D layered materials have been summarized in a recent review by Autere *et al*[19,20].

In this context, we have used confocal spectroscopy to simultaneously measure SHG and THG intensities of exfoliated single- and few-layers $WSe_2$. By studying the harmonics generation as a function of pump power, layer-dependence and spatial position, and correlating the data with atomic force microscopy, we are able to report a robust characterization of third-harmonic $\chi^{(3)}$. We demonstrate that THG is independent on the polarization angle and is proportional to the square of the number of layers. We further show that the magnitude of $WSe_2$ $\chi^{(3)}$ is comparable to that of sulfides of the same TMD family. Even though this type of experiment has been broadly applied to determine $\chi^{(2)}$ and $\chi^{(3)}$ in several two-dimensional materials[14,18,21], the characterization of THG and a quantitative measurement of $\chi^{(3)}$ in $WSe_2$ had remained unknown.

## Results and discussion

We investigated the nonlinear optical properties of mechanically exfoliated atomically thin 2H-WSe$_2$ (2H polytype) with a home-built confocal microscope specifically designed for harmonics generation. Before nonlinear optical experiments, the sample was extensively characterized by photoluminescence and Raman spectroscopy, as well as by atomic force microscopy (for details on the sample preparation, characterization and experimental setup, see materials and methods section and supporting information). The pump wavelength for the harmonic experiments is 1546 nm, therefore SHG emission is centered at 773 nm and THG emission at 516 nm, as shown in Figure 1.

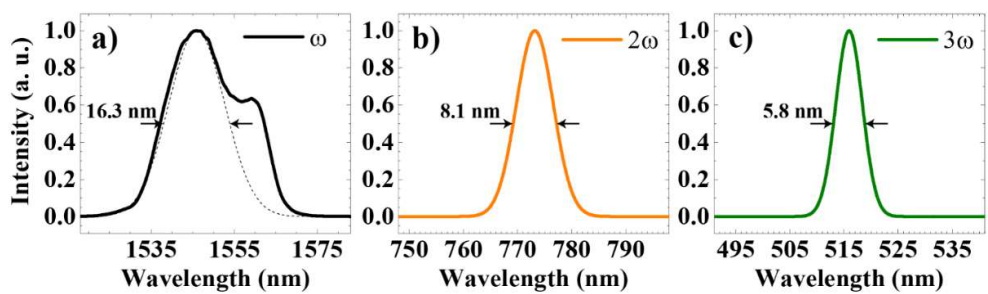

*Figure 1. Spectra for a) 1550 nm pump, b) SHG at $\lambda_{central}$ = 773 nm and c) THG at $\lambda_{central}$ = 516 nm. The pump beam full-width at half-maximum (FWHM) bandwidth is 25.1 nm, but only the peak centered at 1546 nm contributes for the harmonic generation (dotted line, FWHM bandwidth ~ 16.3 nm). The FWHM bandwidth and central wavelength for SHG are 8.1 nm and 773 nm, and for THG are 5.8 nm and 516 nm, respectively.*

The absolute orientation of the WSe$_2$ crystal, relatively to the laboratory frame, was determined by polarized second-harmonic generation (pSHG), where the incident pump linear polarization and a parallel polarizer were rotated simultaneously by the same angle $\theta$ while recording the spectrum. Since 2H-WSe$_2$ belongs to the D3h point group[22], as expected from the threefold rotational symmetry along the c crystallographic axis and demonstrated for a number of odd-layered TMD of this family, the pSHG intensity is proportional to $\cos^2[3(\theta-\phi_0)]$, where $\phi_0$ is the angle of the armchair direction of the flake in the laboratory frame. The result of our pSHG is shown in Figure 2a. This strong polarization dependence makes pSHG a preferred tool for fast crystallographic alignment of these materials[7,8,14,23]. In contrast, the THG is polarization independent, as can be seen from Figure 2b, as expected for all other crystals belonging to the D3h group[18]. Figure 2c shows an optical image of the sample with the identification of armchair (AC) and zig-zag (ZZ) directions of the flakes[8].

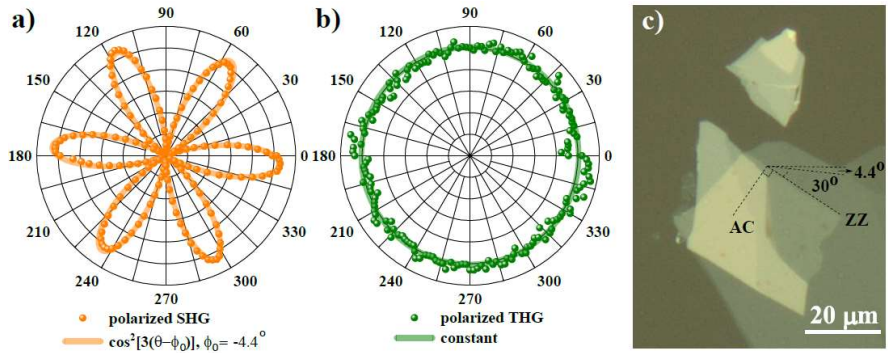

*Figure 2. Polarized harmonic generation in WSe$_2$: a) SHG, $\phi_0$ = -4.4° and b) THG, constant intensity with $\theta$; c) optical image of the sample with armchair (AC) and zig-zag (ZZ) directions.*

We hence determined the AC direction to lie at $\phi_0 = -4.4° + m \times 60°$, $m \in \mathbb{Z}$, and set the pump polarization parallel to this direction of maximum response for our subsequent SHG and THG experiments.

To assert the second and third-harmonic nature of the signals discussed so far, we measured their dependence on the pump power via Malus' law experiment, by inserting a fixed polarizer before the sample in the experimental setup used for pSHG measurements (for more details on the experimental setup, see supporting information). Additionally, we carried out a calibration leading to actual average power readings from spectrometer counts in order to extract the magnitude of the susceptibilities (for more details on calibration factors, see supporting information). Figure 3 shows power-dependent double logarithmic plots of SHG and THG actual average power. The power scaling of the SHG and THG intensities is shown to follow very well the respectively expected quadratic and cubic dependences.

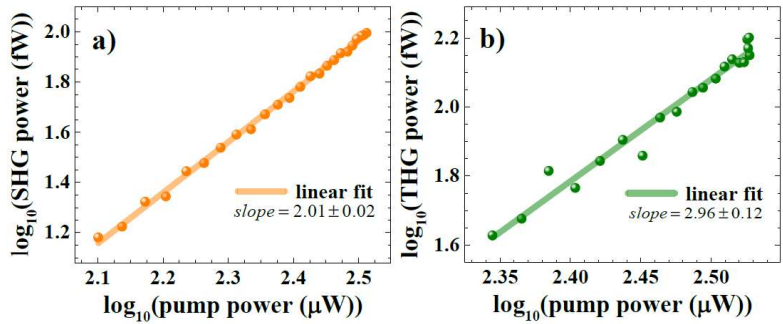

*Figure 3. Double logarithmic plots of the power-dependent nonlinear signals for a) the SHG of single-layer and b) the THG of 9-layers WSe$_2$.*

To acquire spatial maps, the sample was raster scanned across the pump beam in 0.5 μm steps, while SHG and THG intensities were recorded simultaneously. The results are shown in Figure 4. An image of the sample with the number of layers (*N*) labeled in each region, is shown in Figure 4a. The thickness assignment was performed by correlating atomic force microscopy data with the SHG contrast. The image also shows the presence of a thin film of hexagonal boron nitride (hBN) partially covering the WSe$_2$ multilayer, which was used to prevent environmental degradation (for more details on sample fabrication, see supporting information).

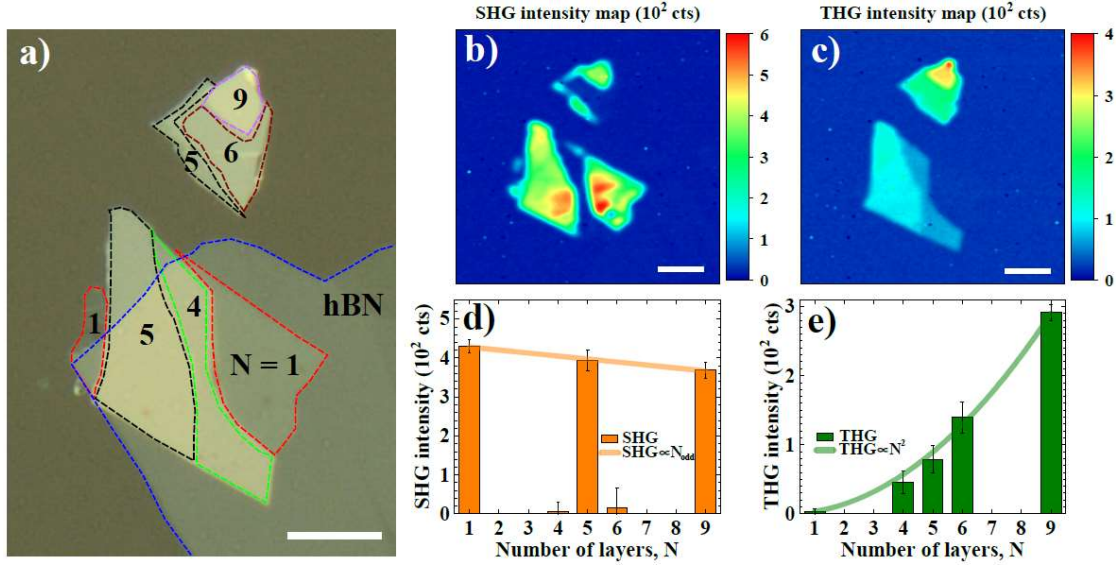

*Figure 4. a) Optical image of the sample with labeled number of layers (N); b) Spatial SHG and c) THG intensity mappings across the WSe$_2$ sample; d) Spatial average SHG and e) THG intensities as a function of N. The values and error bars indicated for the SHG and THG correspond to the mean ±3×(standard deviation). Scale bars = 10 μm.*

Figure 4b shows the sharp SHG contrast between regions of odd- and even-*N*, as SHG in these TMD is expected for odd-*N* flakes[24,25], with a decreasing intensity for increasing number of layers[26,27]. Yet, even though SHG is expected to be strictly zero for even-*N*, a residual non-zero intensity is recorded for 4- and 6-layers regions, as shown in Figure 4d. This effect might be attributed to incomplete destructive interference between SHG from adjacent layers, as discussed previously by Li *et al*[25], although its fundamental cause remains unexplored. Moreover, this effect might be responsible for a small negative slope for the SHG intensity as a function of the odd-*N*. Extrapolating the linearly decreasing trend inferred from Figure 4d, significant SHG intensity in WSe$_2$ should be observed for samples with up to ~50 layers. However, the challenges in unambiguously determining *N* for samples with more than 10 layers make this observation difficult.

Figure 4c shows the THG map of the sample. In addition to the signal now arising from all regions of the sample, the key observation is that the intensity clearly scales up with the number of layers. To be more specific and quantitative, we compiled the average intensity over regions with same *N* in Figure 4c. The THG intensity is proportional to $N^2$, as shown in Figure 4e. This quadratic scaling with thickness is a direct evidence that we can consider each layer as contributing independently to the overall THG, since the linear relation of the third-harmonic susceptibility $|\chi^{(3)}{}_N|$ = $N \times |\chi^{(3)}{}_s|$ indeed implies in $I_{THG} \propto |\chi^{(3)}{}_N|^2 \propto N^2 \times |\chi^{(3)}{}_s|^2$.

No border effect for enhancement of SHG or THG was observed. In fact, DFT calculations for chemical vapour deposited (CVD) $WS_2$ showed that at the borders the bandgap becomes indirect[28,29], which actually contributes to quenching of SHG and, probably, THG at the edges.

A direct comparison between Figure 4a and Figure 4c indicates that regions with same $N$ have slightly weaker harmonic intensities when covered by hBN, which the most notable case being the 5-layers region. This can be explained by the Fresnel reflection/transmission coefficients, since the total transmittance depends on the layer stacking, total sample thickness and refractive index mismatch. We observed that, by recording the transmitted power at the fundamental frequency with a reference photodetector, regions without hBN may have up to 4% higher pump transmittance than covered regions (for more details on reference transmittance mapping, see supporting information). This difference in pump power leads to SHG and THG intensities of, respectively, 8% and 12% higher in non-covered regions. The analysis presented in Figure 4d and Figure 4e already corrects the intensities by the pump power in each region. The contribution to SHG and THG from hBN solely is negligible, as we have confirmed by pumping hBN regions of the sample (Figure 4b and Figure 4c). This agrees with previous report by Li *et al.*[25], where the authors show that hBN $|\chi^{(2)}{}_s|$ is 2 orders of magnitude lower than that of TMD.

Taking advantage of the spatial resolution and well defined layer assignment in our experiment, the magnitude of the nonlinear susceptibilities $\chi^{(2)}$ and $\chi^{(3)}$ for $WSe_2$ were extracted from the harmonic mapping by using the model deducted by Woodward *et al*[18]. Simple modifications were implemented in the model, accounting for harmonic generation in transmittance, with fundamental beam pumping the sample from the substrate-sample interface, as shown in Equation (1) and (2):

$$\left|\chi_s^{(2)}\right| = \sqrt{\frac{P(2\omega)}{P^2(\omega)}} \times \frac{c.\varepsilon_0.RR.A.\Delta\tau.\lambda^2}{64\sqrt{2}\pi^2 S} \times \frac{(1+n)^6}{n^3}, \tag{1}$$

$$\left|\chi_s^{(3)}\right| = \sqrt{\frac{P(3\omega)}{P^3(\omega)}} \times \frac{c^2.\varepsilon_0^2.RR^2.A^2.\Delta\tau^2.\lambda^2}{336\sqrt{3}\pi^2 S^2} \times \frac{(1+n)^8}{n^4}, \tag{2}$$

where $P(2\omega)$ and $P(3\omega)$ are second-harmonic and third-harmonic average powers, respectively, and $P(\omega)$ is the fundamental average power. $c$ is the speed of light in vacuum, $\varepsilon_0$ is the vacuum electric permittivity. $RR$ is the repetition rate (80 MHz), $A$ is the minimum spot area (2.0 ± 0.3 µm², obtained from the half-width at $1/e^2$ beam radius of 0.8 ± 0.1 µm), $\Delta\tau$ is the full-width at half maximum pulse duration at the sample spot (200 ± 10 fs) and $\lambda$ is the wavelength (1545 nm) for the fundamental beam. Finally, $n$ is the substrate refractive index at the fundamental wavelength (1.47 at 1545 nm) and S = 0.94 is a shape factor assuming Gaussian pulses. The average power of the fundamental beam was kept constant at 0.65 mW, and the calibration factors to obtain harmonic average powers from counts, as in Figure 4, are 0.275 fW counts$^{-1}$ and 0.512 fW counts$^{-1}$, respectively, for SHG and THG. $|\chi^{(2)}|$ and $|\chi^{(3)}|$ results are summarized in Table 1.

Table 1. Harmonic susceptibilities of WSe$_2$ as a function of number of layers N. The reported values correspond to the mean ± standard deviation. The error values are obtained from the uncertainties of the relevant experimental parameters present in equations (1) and (2), through classical error propagation theory.

| Number of layers (N) | $|\chi^{(2)}_N|$ ($10^{-19}$ m² V$^{-1}$) | $|\chi^{(3)}_N|$ ($10^{-28}$ m³ V$^{-2}$) |
|---|---|---|
| 1 | 0.70 ± 0.09 | 0.9 ± 0.2 |
| 4 | 0.08 ± 0.08 | 3.4 ± 0.5 |
| 5 | 0.67 ± 0.09 | 4.5 ± 0.6 |
| 6 | 0.2 ± 0.1 | 6.0 ± 0.8 |
| 9 | 0.65 ± 0.08 | 8.6 ± 0.9 |
| Average/sheet (per layer) | - | 0.91 ± 0.07 |
|  | $|\chi^{(2)}_b|$ ($10^{-10}$ m V$^{-1}$) | $|\chi^{(3)}_b|$ ($10^{-19}$ m² V$^{-2}$) |
| Bulk susceptibility | 0.9 ± 0.1 | 1.16 ± 0.09 |

As only odd-$N$ regions contribute with appreciable SHG, the second-order susceptibility is expressed in terms of an effective value $|\chi^{(2)}_N|$, which was directly obtained from each region on the WSe$_2$ sample. The third-order susceptibility is expressed in terms of both an effective value, $|\chi^{(3)}_N|$, and the average per layer, $|\chi^{(3)}_s|$. The bulk-like values are obtained from the expression $|\chi_b| = |\chi_s|/\delta$, where $\delta$ is the inter-layer distance of the flake (for our WSe$_2$ flake, $\delta$ = 0.79 ± 0.02 nm – for more details on the thickness characterization, see supporting information).

The values obtained for sheet and bulk second-order susceptibility agree with those previously reported[7,30] within the order of magnitude, and the sheet and bulk values of third-harmonic susceptibility are reported here for the first time. The $\chi^{(3)}$ magnitude is comparable to other TMD[18,31] and is larger than that reported for graphene[18,32]. Although the presented results have small errors inherent from the model and parameters we used (~13% for SHG at odd-N, ~13% for THG – except N=1 with 20% error), those should only be taken as a good estimative for the order of magnitude of such susceptibilities, since their determination relies on the precise measurement of many important parameters (which can vary upon definitions and measurement techniques). The measurement of harmonic susceptibilities of two-dimensional layered materials is known to be sensitive to the fabrication process (Woodward et al.[18] reported 26% variation between exfoliated and CVD $MoS_2$), stacking order, and surrounding environment (substrate and superstrate)[19].

Excitons are strongly present in semiconductor TMD, excitonic effects may also play role by enhancing nonlinear optical transitions, as reported previously for $MoS_2$[8] and for $WSe_2$[33,34]: when the harmonic photon energies are in resonances to excitonic or single-particle energy levels from the material, harmonic signals up to 1 order of magnitude higher might be observed. Since in our experiment $WSe_2$ is pumped by photons of 0.8 eV, no resonant enhancement effect is expected to take place, because both SHG (1.6 eV) and THG (2.4 eV) are off-resonance with energy levels of the material[35]. Nevertheless, our SHG results agrees to those off-resonance published results. Although no resonance has been observed, the presented characterization provides important parameters to support the realization of 2D-materials-based devices for applications in telecommunication systems and silicon photonics.

Benefiting from the small sample thickness and THG efficiency, multi-layer $WSe_2$ appears to be a potential material for nonlinear optical applications as, for example, on-chip optical frequency conversion[36], silicon photonics[37,38], or other third-order related phenomena like all-optical switching[39,40], which depends on a different $\chi^{(3)}$.

## Conclusion

By extensively characterizing single- and few-layers flakes of 2H-$WSe_2$ with photoluminescence, Raman spectroscopy and atomic force microscopy, we obtained the precise number of layers of each regions of the flake. We map the spatial emission of the second- and third-harmonic signals to study the layer dependence of the nonlinear response and quantify the corresponding susceptibilities. We obtained the values $|\chi^{(3)}_s| = (0.91 \pm 0.07) \times 10^{-28}$ m³ V⁻² and $|\chi^{(3)}_b| = (1.16 \pm 0.09) \times 10^{-19}$ m² V⁻², which are comparable to the third-harmonic susceptibility of related semiconducting TMD and provide one step forward towards the complete characterization of the nonlinear optical properties of this family ($MX_2$; M = W, Mo; X = S, Se). The reliability of these values is supported by the good agreement of our values for $|\chi^{(2)}|$, extracted from the simultaneously recorded second-harmonic signal, with previous reports for the same material. The strong nonlinear response of WSe$_2$ in the infrared makes it a material suitable for applications in silicon photonics, all-optical switching and optical frequency conversion.

## Materials and methods

**Sample fabrication and characterization**: flakes of WSe$_2$ were obtained via micromechanical exfoliation from a single bulk crystal and transferred onto a fused silica substrate. To prevent degradation, the sample was partially covered with multi-layer hBN. The sample's properties were characterized by photoluminescence, Raman spectroscopy and atomic force microscopy (for more details sample fabrication and characterization, see supporting information).

**Experimental setup**: Nonlinear optical properties of WSe$_2$ were investigated in a home-built confocal multiphoton microscope setup. We used a half-waveplate to control the input linear polarization and a polarizer to analyze the harmonic signals. The pump laser was a 1545 nm, 200 fs, 80 MHz mode-locked fiber. A 100x objective lens focused the pump beam down to a spot size of 2 µm$^2$. The step used in the sample displacement during spatial mapping was 0.5 µm (for more details on the experimental setup, see supporting information).

## Acknowledgements

H. Y. W. acknowledges scholarship support from NGS. G.E. acknowledges financial support from National Research Foundation of Singapore (NRF Research Fellowship NRF-NRFF2011-02 and medium-sized centre programme) and Ministry of Education of Singapore (AcRF Tier 2 MOE2015-T2-2-123). V. M. P. acknowledges financial support from Ministry of Education of Singapore (FRC AcRF Tier 1 R-144-000-386-114). J. C. V. G. acknowledges financial support from CA2DM through National Research Foundation of Singapore (NRF-CRP Grant No. R-144-000-295-281).

## Author contributions

H. G. R. and H. Y. W. performed sample characterization, nonlinear optical experiments and analyzed the data. I. V. fabricated the sample. M. J. F. L. R set up the confocal microscope system. T. T. and K. W. provided the hexagonal boron nitride sample. G. E., V. M. P. and J. C. V. G. supervised the work and revised the manuscript. All authors discussed and interpreted the results and contributed to the writing of the manuscript.

## Additional Information

**Competing interests**: all authors declare no competing interests.

**Supporting information:** sample characterization with optical microscopy, photoluminescence microscopy, Raman spectroscopy, atomic force microscopy, nonlinear optics experimental setup configuration.

## Data Availability

The datasets generated during and/or analyzed during the current study are available from the corresponding author on reasonable request.

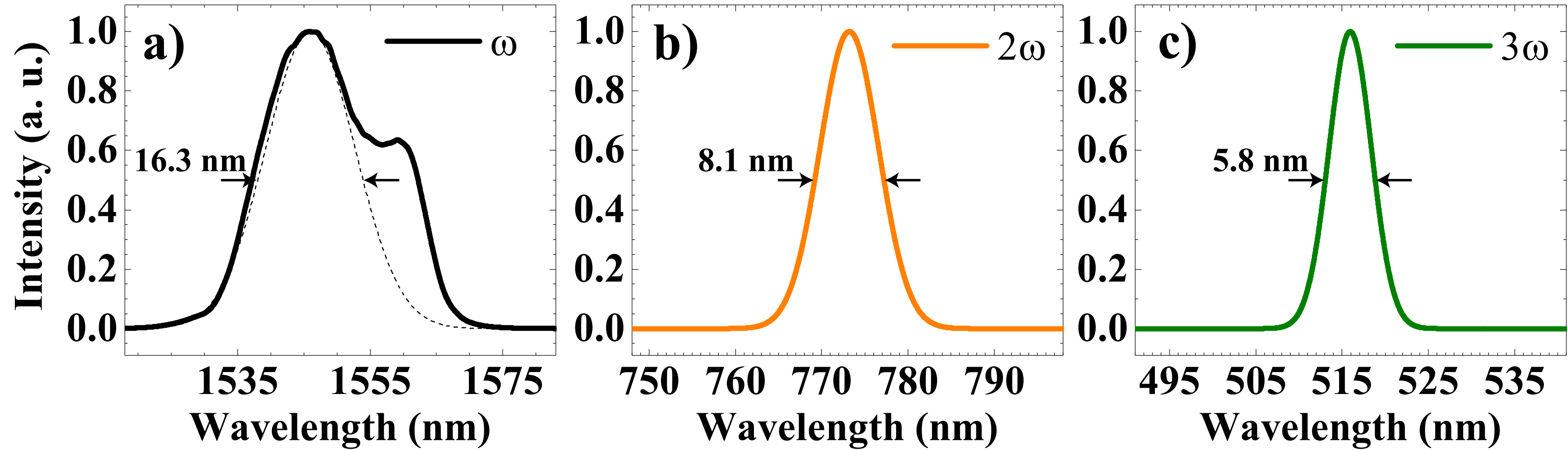

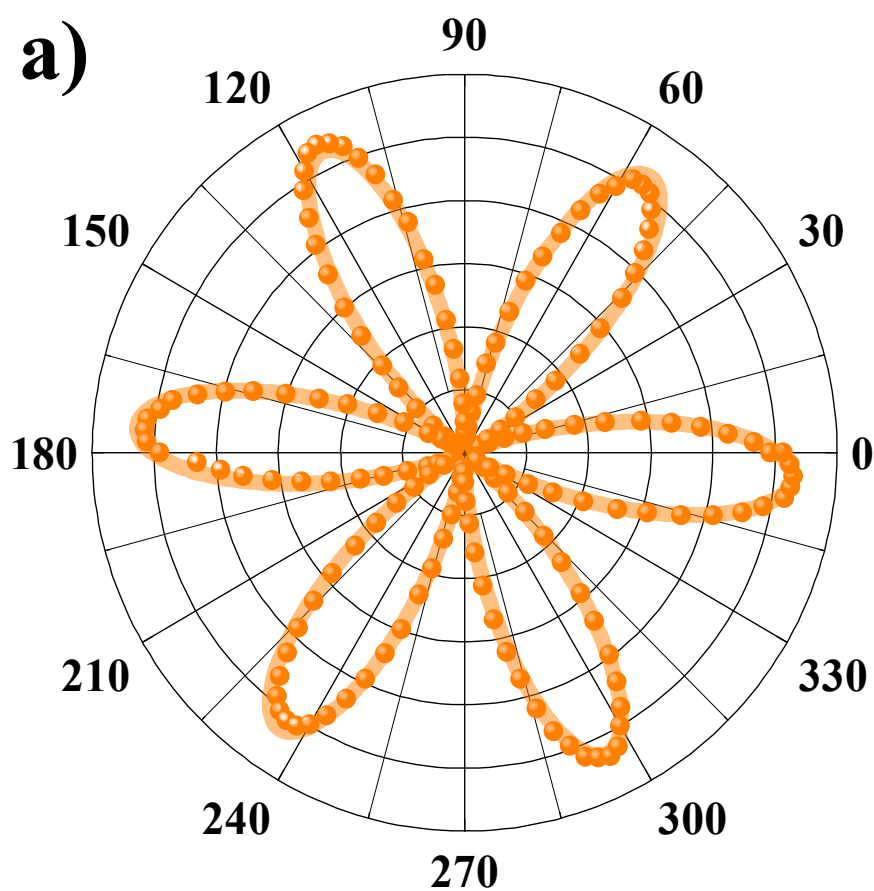 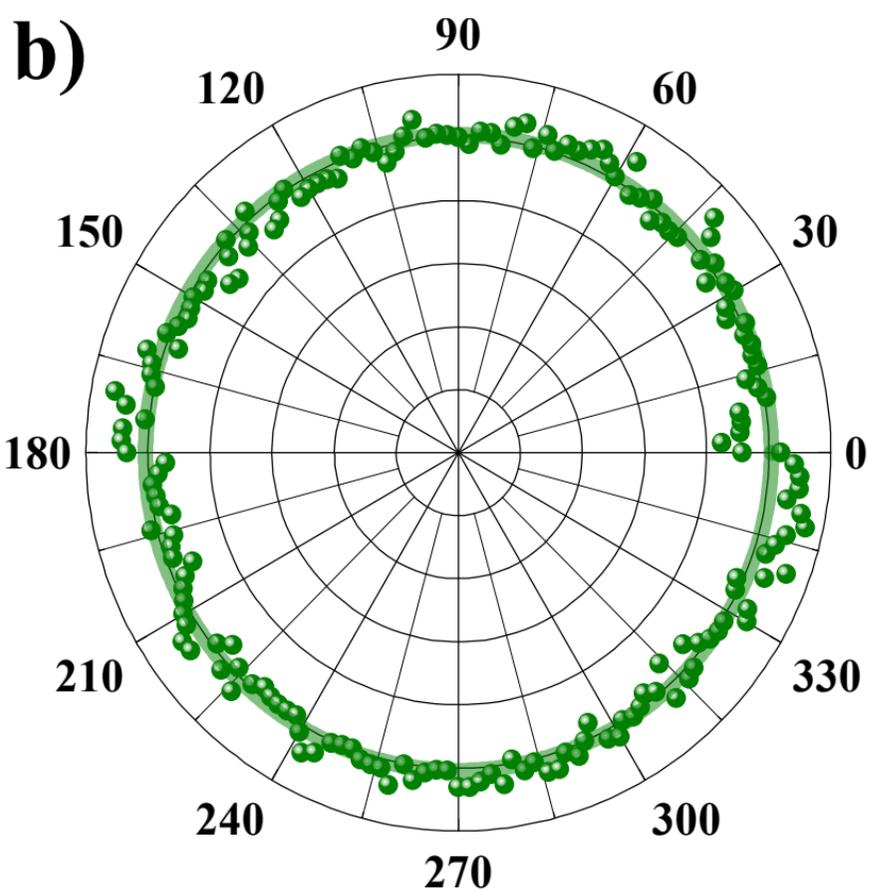 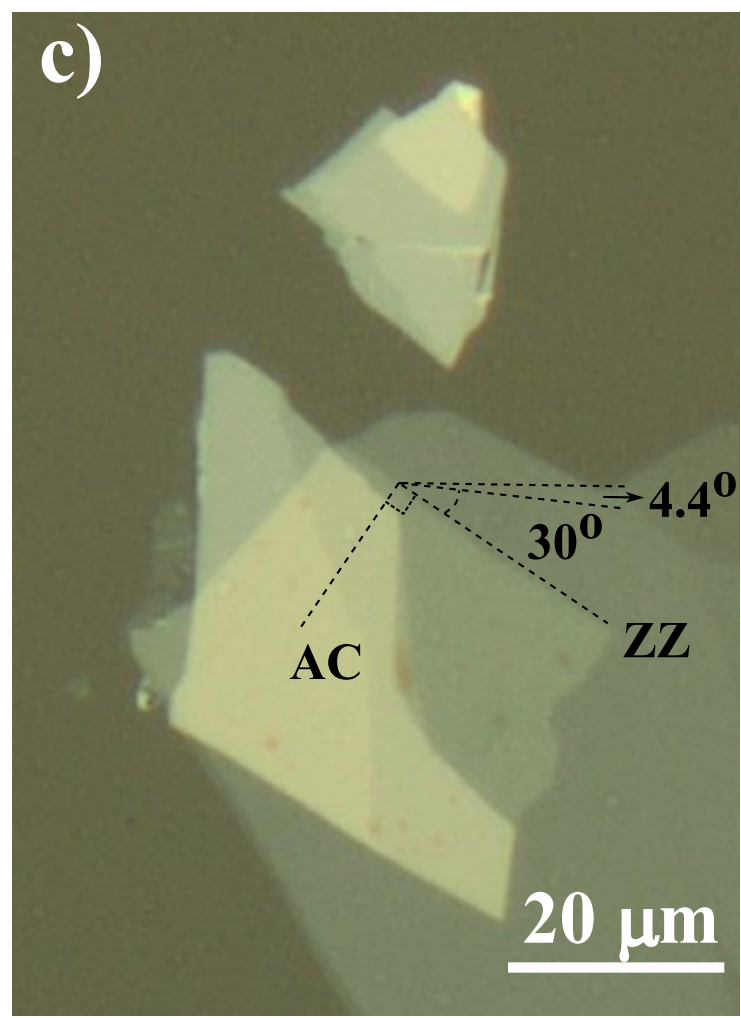

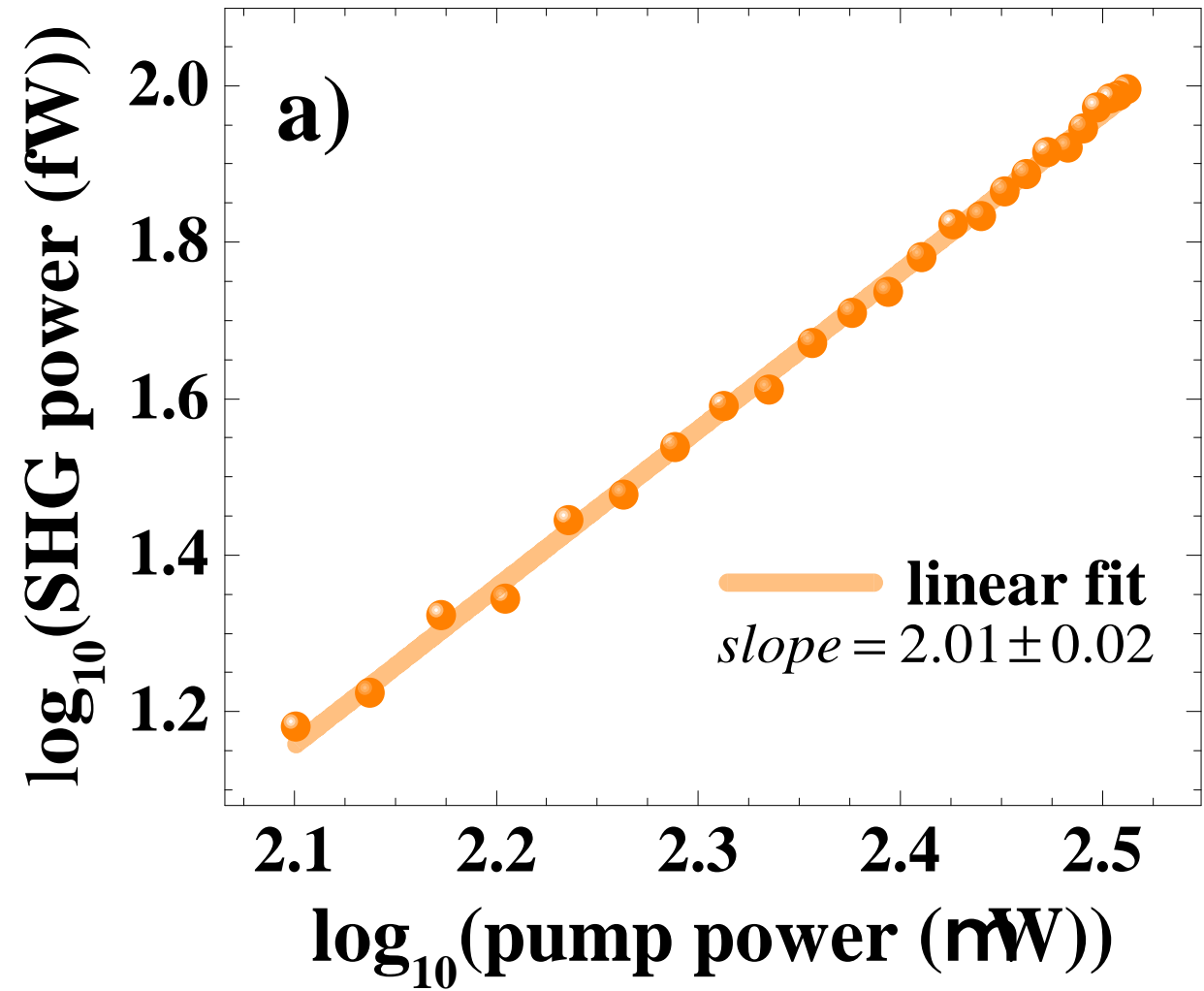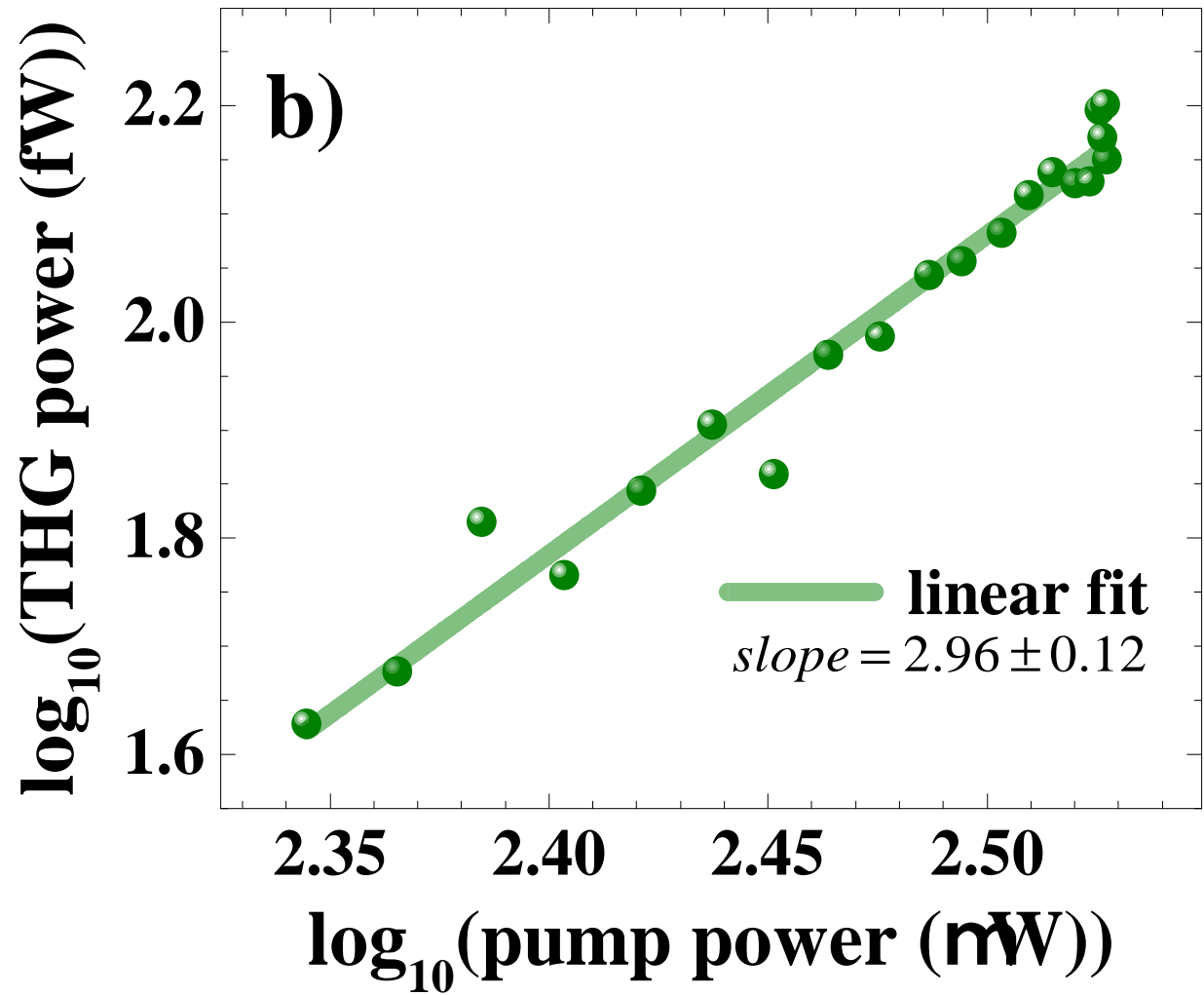

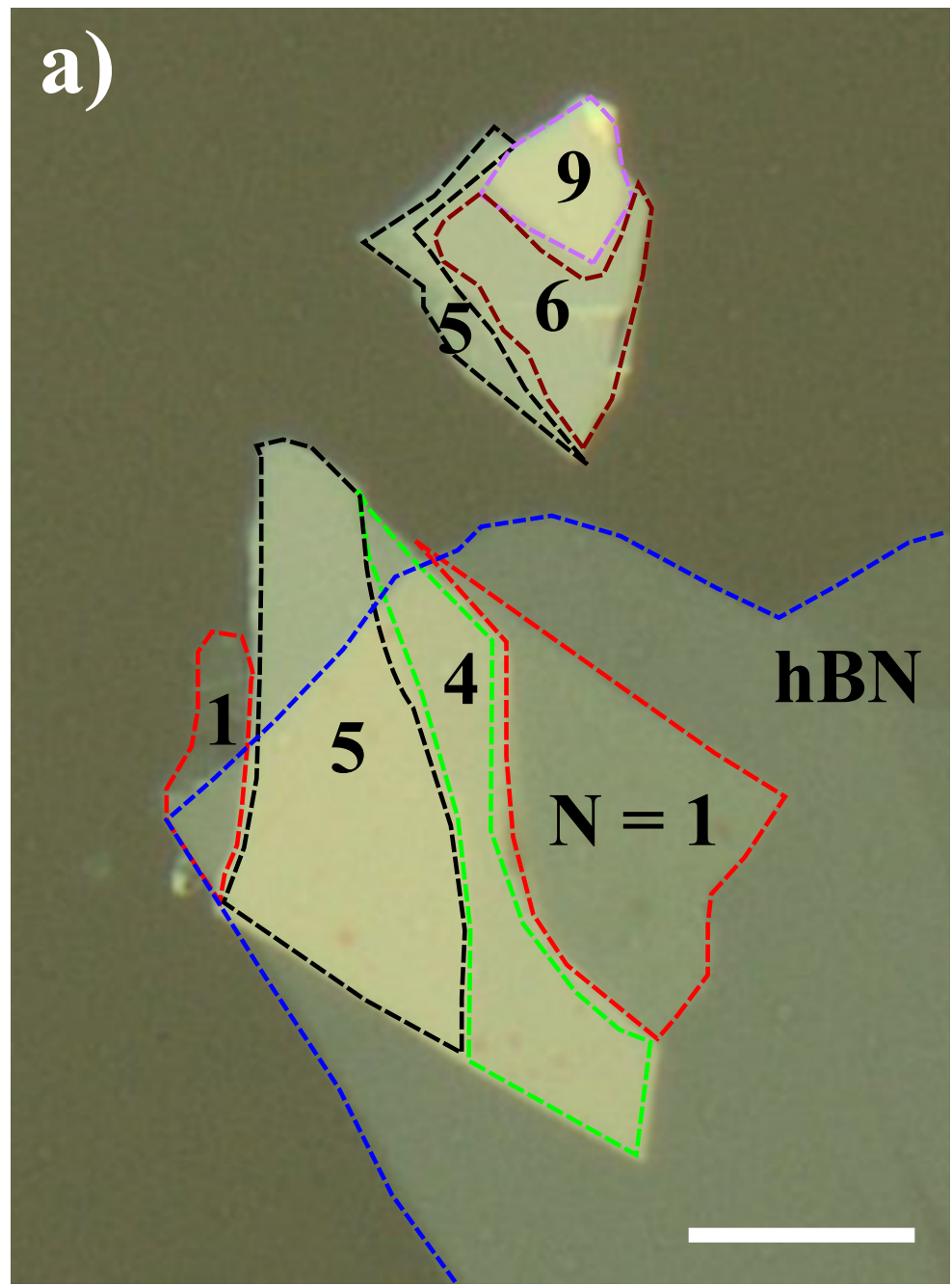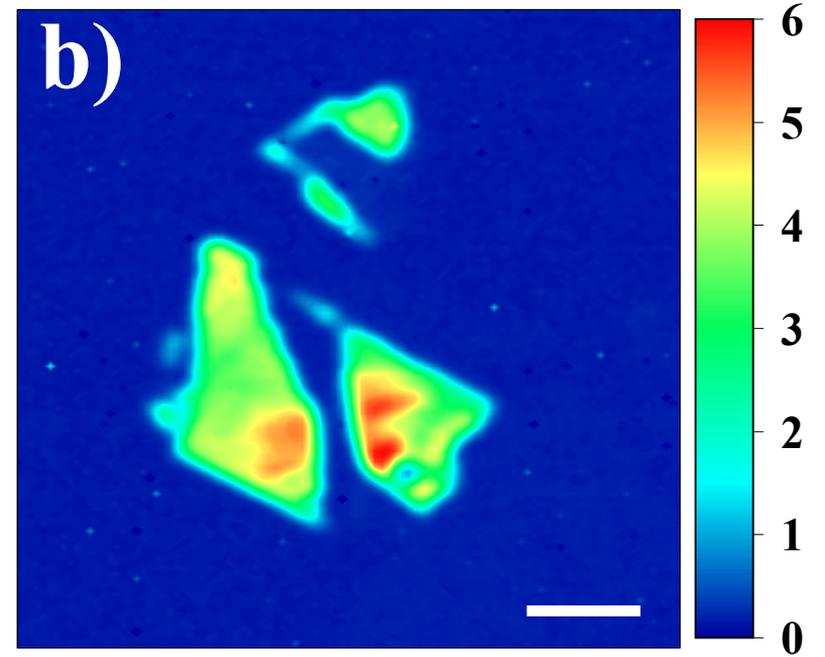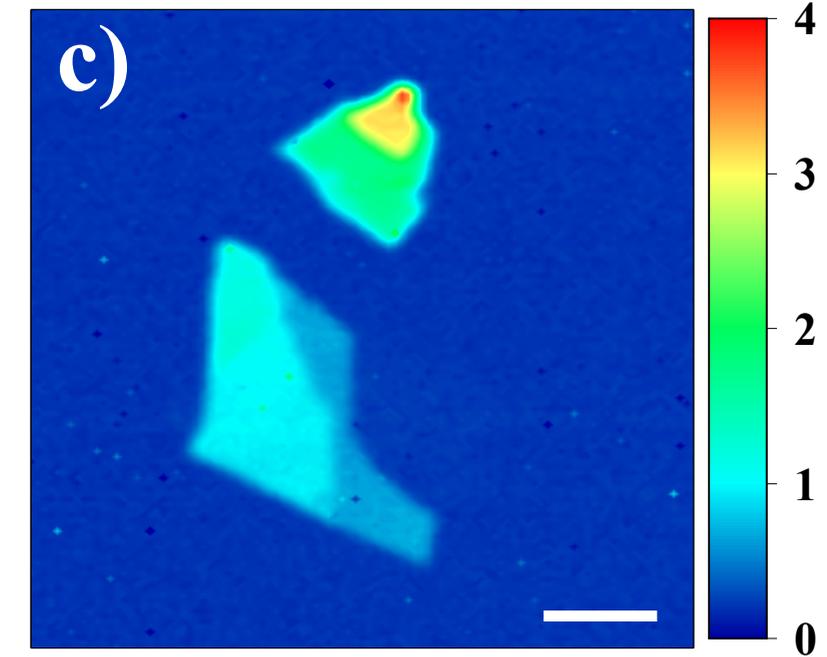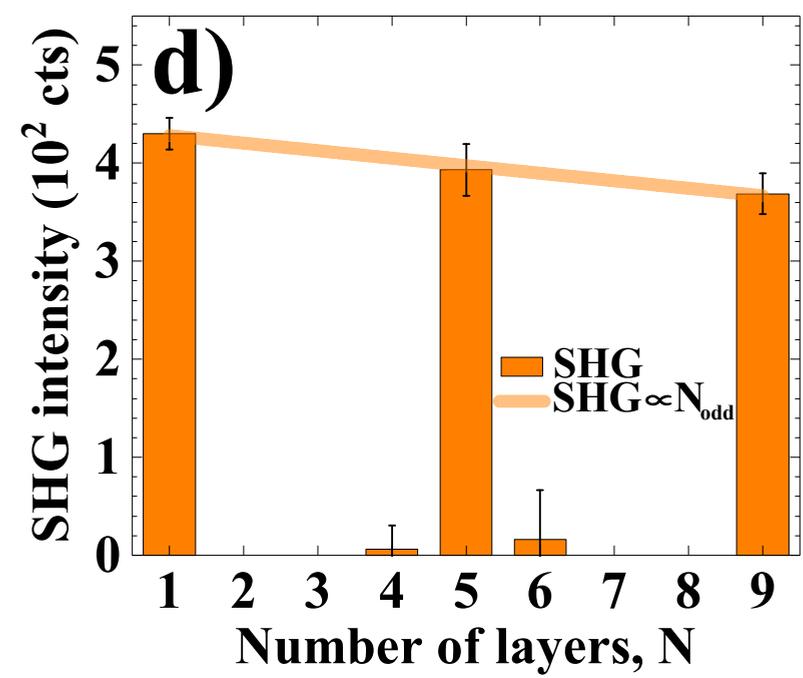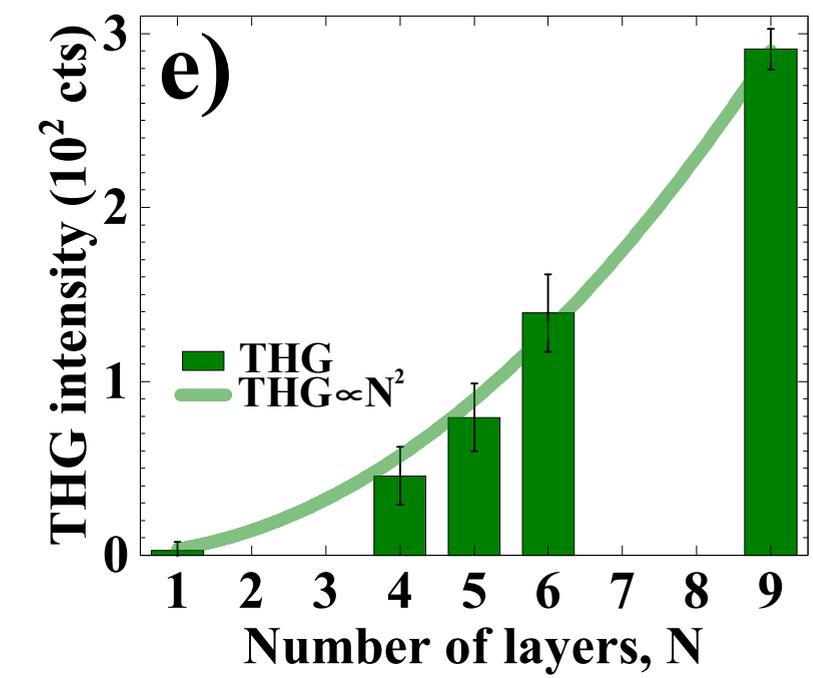

# Characterization of the second- and third-harmonic optical susceptibilities of atomically thin tungsten diselenide

Henrique G. Rosa[1,*], Ho Yi Wei[1,2,3,*], Ivan Verzhbitskiy[1,3], Manuel J. F. L. Rodrigues[1,4], Takashi Taniguchi[5], Kenji Watanabe[5], Goki Eda[1,3,6], Vitor M. Pereira[1,3, †], José C. V. Gomes[1,3, ‡]

[1]Centre for Advanced 2D Materials, National University of Singapore, 6 Science Drive 2, Singapore 117546

[2]NUS Graduate School for Integrative Sciences and Engineering (NGS), Centre for Life Sciences (CeLS), 28 Medical Drive, Singapore 117456

[3]Department of Physics, National University of Singapore, 2 Science Drive 3, Singapore 117551

[4]Center of Physics and Department of Physics, Universidade do Minho, 4710-057, Braga, Portugal

[5]National Institute for Materials Science, 1-1 Namiki, Tsukuba 305-0044, Japan

[6]Department of Chemistry, National University of Singapore, 3 Science Drive 3, Singapore 117543

*These authors contributed equally to this work

†Corresponding author email: vpereira@nus.edu.sg

‡Corresponding author email: phyvjc@nus.edu.sg

## Supporting information

**Sample fabrication:** We fabricated the tungsten diselenide ($WSe_2$) sample via micromechanical exfoliation from a bulk crystal and directly transferred onto glass substrate from the exfoliation tape. Then, multi-layer hexagonal boron nitride (hBN), approximately 16 nm thick, was transferred onto $WSe_2$, partially covering the flake to protect it from environmental degradation[1,2].

**Sample characterization:** We used photoluminescence (PL) microscopy and Raman spectroscopy to characterize sample's properties and to identify monolayer regions, as shown in Figure S1.

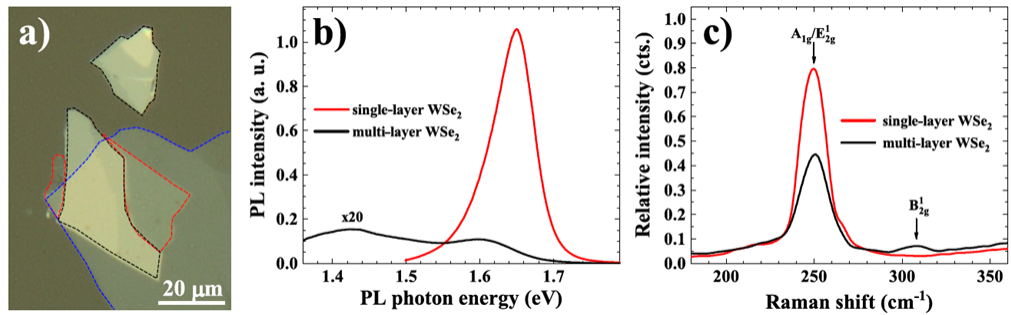

*Figure S1. a) Optical image of WSe₂/hBN structure - dashed red line: single-layer regions, dashed black line: multi-layer regions, dashed blue line: multi-layer hBN; b) photoluminescence and c) Raman spectra of single- and multi-layer regions.*

From both PL and Raman spectra shown in Figure S1, one can distinguish between single- and multi-layer regions within the sample: single-layer region presents a prominent PL peak at 1.65 eV and degenerate $A_{1g}/E_{2g}^1$ Raman peak at 249.6 cm$^{-1}$[3]. For all multi-layer regions, besides the flat and weak PL signal, the $B_{2g}^1$ Raman peak (characteristic of few-layers WSe₂)[3] is present. As expected, no PL or Raman peak associated with hBN were observed.

To unambiguously determine the precise number of layers (*N*) of each region, atomic force microscopy (AFM) was used to trace profile cuts from substrate onto each region of the sample. Figure S2 shows the AFM results.

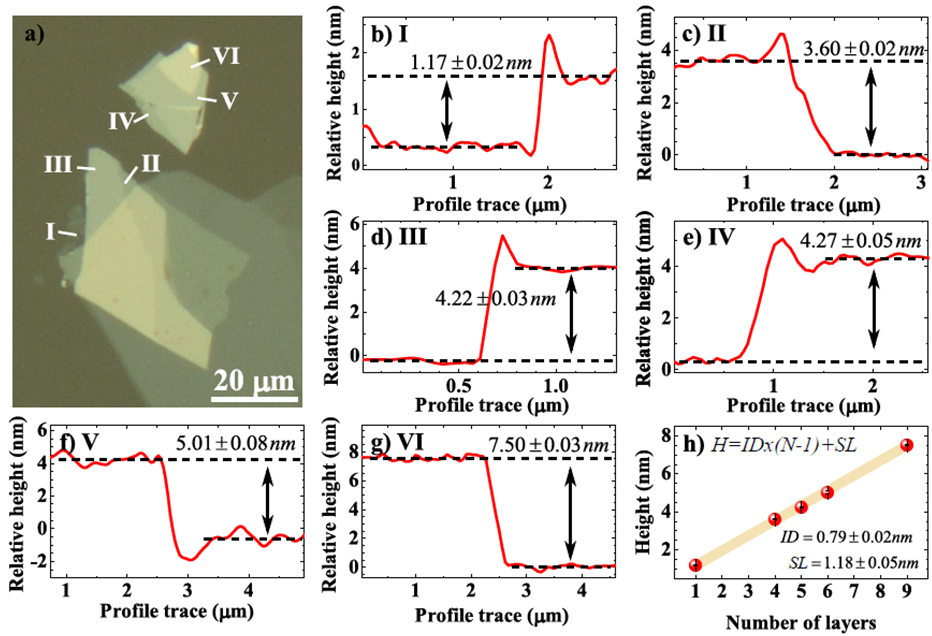

*Figure S2. a) Sample optical image indicating the AFM profile traces location; b-g) profile traces I to VI; c) Measured height for each region, allowing the identification of the number of layers – ID: inter-layer distance, SL: single-layer thickness.*

From PL and Raman results, trace I (Figure S2b) shows the thickness for the single-layer region to be ~1.17 nm, which agrees with previous published results[4–6]. The inter-layer separation distance is extracted from traces II – VI (Figure S2c-g), found to be in the 0.7-0.8 nm range. A linear fit (Figure S2h) yields single-layer thickness and inter-layer distance of 1.18 ± 0.05 nm and 0.79 ± 0.02 nm, respectively. To summarize, regions I to VI were shown to have 1, 4, 5, 5, 6 and 9 layers, respectively.

**Experimental setup:** Nonlinear optical properties of $WSe_2$ were investigated in a home-built confocal multiphoton microscope setup, as shown in Figure S3.

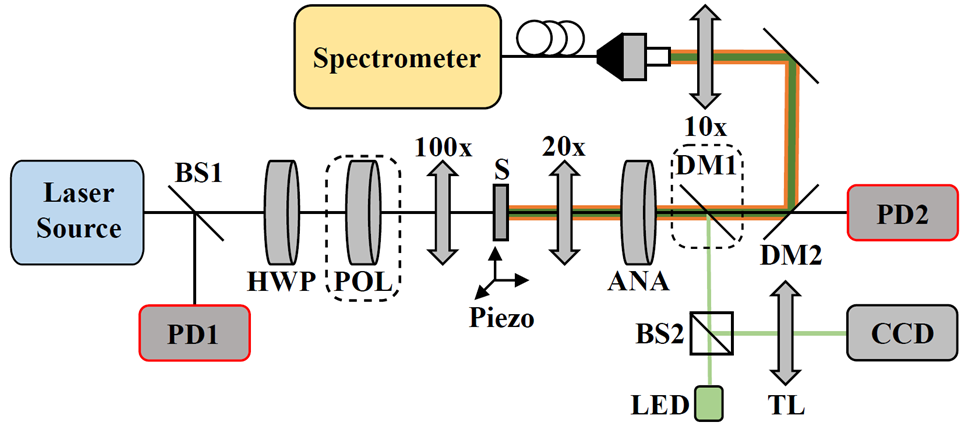

*Figure S3. Confocal multiphoton microscope experimental setup. BS: beam splitter; PD: photodetector; HWP: half-waveplate; POL: polarizer; 100x: focusing 100x objective lens; S: sample; 20x: imaging 20x objective lens; ANA: analyzer; DM: dichroic mirror; TL: tube lens; CCD: CCD imaging camera; 10x: coupling 10x objective lens.*

The microscope setup has two parts: the first part is an optical microscope for locating and positioning the sample. For this purpose, DM1 is placed in the setup, allowing sample's illumination and back-reflected light collection. A piezoelectric stage is docked on a mechanical translational stage to control sample's position precisely. The sample is mounted on the setup such that the $WSe_2$ flake faces the 20x objective.

The second part is the confocal multiphoton microscope for SHG and THG measurements. In this experiment, a 1550 nm, 200 fs, 80 MHz mode-locked fiber laser was used as pump beam. The 100x objective lens focused the pump beam down to a spot size of 2 $\mu m^2$. The average power of the pump, at the sample, was 0.65 mW. The sample was mounted on the setup such that the $WSe_2$ flake faces the 20x objective. After the sample, pump and harmonic beams were collimated by a 20x objective lens, then separated by DM2. The pump was recorded by a reference photodetector (PD2) while harmonic signals were coupled to an optical fiber by a 10x objective lens and detected by a TEC-CCD spectrometer (Avantes HERO) with 10 s acquisition time. All characterization and experiments were performed at room temperature.

As can be seen from Figure S1, the optical bandgap energy of the single-layer flake is 1.65 eV, which places both second- and third-harmonic signals off resonance, therefore no signal enhancement caused by excitonic effect was observed.

The calibration factors for the actual power reading by the spectrometer were determined by simultaneously measuring the power of a reference laser beam in a powermeter and in the spectrometer (with calibrated attenuators). The calibration factor for 775 nm (SHG) is 0.275 fW counts$^{-1}$ and for 516 nm (THG) is 0.512 fW counts$^{-1}$, accounting for the direct relation of generated harmonics power at the sample and the spectrometer reading.

**Transmitted pump power mapping:** We recorded simultaneously the transmitted pump power for each position of the harmonic mappings, therefore being able to also map the WSe$_2$ sample in transmittance at the fundamental wavelength. The pump beam map is shown in Figure S4. From this map we can observe that different regions of the sample, depending upon the WSe$_2$-hBN stacking structure, have different transmitted pump powers. This can be attributed to the different stacking at each region or even two-photon absorption. We did not investigate further for the latter effect as it is not the focus of this report.

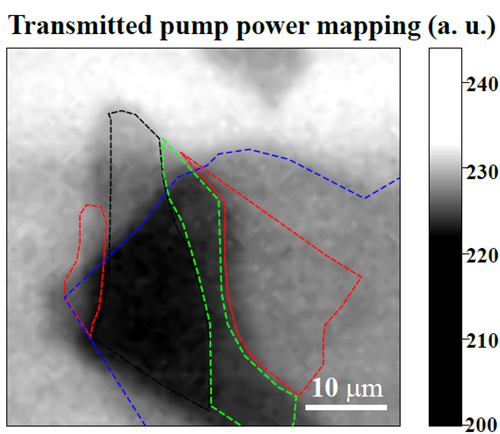

*Figure S4. Pump beam transmittance mapping.*